\documentclass[twocolumn]{jpsj3}
\usepackage[usenames]{color}
\def\v#1{\mib #1}
\newcommand{\bra}[1]{\left\langle {#1} \right\vert}
\newcommand{\ket}[1]{\left\vert {#1} \right\rangle}

\newcommand{\aver}[1]{\left\langle {#1} \right\rangle}
\newcommand{\avercl}[1]{\left\langle {#1}_{\rm cl}\right\rangle}

\def\Jeff{J_{\rm eff}}
\def\Deff{D_{\rm eff}}
\def\dfrac#1#2{{\displaystyle\frac{#1}{#2}}}
\def\scl{M_{\rm cl}}
\def\Ecl{E_{\rm cl}}
\def\EHal{{\tilde{E}}}
\def\epsgtilde{{\tilde{\epsilon}_{\rm G}}}
\def\theenumi{{\roman{enumi}}} 
\def\runauthor{K. {\sc Hida} and K. {\sc Takano}}
\title
{Effects of Single-site Anisotropy on Mixed Diamond Chains with Spins $1$ and $1/2$}

\author
{Kazuo {\sc Hida}\thanks{E-mail address: hida@mail.saitama-u.ac.jp} 
and Ken'ichi {\sc Takano}$^{1}$ 
}

\inst
{Division of Material Science, Graduate School of Science and Engineering, \\ Saitama University, Saitama, Saitama, 338-8570\\ 
$^{1}$Toyota Technological Institute, 
Tenpaku-ku, Nagoya 468-851}

\recdate
{
June 20, 2011
}

\abst{
Effects of single-site anisotropy on mixed diamond chains with spins 1 and
1/2 are investigated in the ground states and at finite temperatures. 
There are phases where the ground state is a spin cluster solid, i.e., an array of uncorrelated spin-1 clusters separated by singlet dimers. 
The ground state is nonmagnetic for the easy-plane anisotropy, while it is paramagnetic for the easy-axis anisotropy. 
Also, there are
 the N\'eel, Haldane, and large-$D$ phases, 
where the ground state is a single spin cluster of infinite size and the system is equivalent to the spin-1 Heisenberg chain with alternating anisotropy. 
The longitudinal and transverse susceptibilities and entropy are calculated at finite temperatures in the spin-cluster-solid phases. 
Their low-temperature behaviors are sensitive to anisotropy. 
}

\kword
{mixed diamond chain, anisotropy, frustration, spin cluster solid, N\'eel phase, Haldane phase, large-$D$ phase
}

\begin{document}
\sloppy
\maketitle
\section{Introduction}

Frustration in quantum spin systems has been one of the most exciting subjects studied in the field of magnetism over these past decades.\cite{diep,intfrust} 
A number of theoretical and experimental investigations have revealed exotic phenomena induced by the interplay of frustration and quantum fluctuation. 
In the theoretical approach, many exactly solvable models have played  crucial roles in elucidating 
frustrated quantum magnetism.
As typical examples, there exists a class of models whose ground states are   
exact spin cluster solid (SCS) states because of frustration. 
The SCS state is defined as a tensor product state of exact local eigenstates of cluster spins. 
A dimer state is the simplest type of  SCS state. 
For example, 
the Majumdar-Ghosh model\cite{mg} has a dimer 
ground state, which is 
a prototype of spontaneously dimerized states 
 in one-dimensional frustrated magnets\cite{hase}. 
The Shastry-Sutherland model\cite{shs}, which 
corresponds to the material 
SrCu$_2$(BO$_3$)$_2$\cite{kage1,kage2}, 
also has a dimer ground state.

\begin{figure} 
%\centerline{\includegraphics[width=4.5cm]{lattice_dia.eps}}
\centerline{\includegraphics[width=4.5cm]{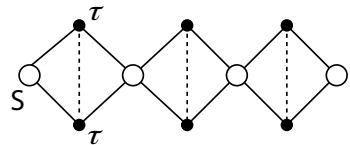}}
\caption{
Structure of the 
diamond chain. 
Spin magnitudes in a unit cell are indicated by $S$ and $\tau$; 
we denote the set of magnitudes by ($S$, $\tau$). 
The PDC is the case of $S = \tau$, while the MDC is the case of $S = 2\tau$ with integer or half-odd integer $\tau$.}
\label{lattice_structure}
\end{figure}

The diamond chain is a
frustrated spin chain with exact SCS ground states 
 that are different from dimer states. 
The lattice structure of the diamond chain 
is shown in Fig.~\ref{lattice_structure}. 
In a unit cell, there are two kinds of nonequivalent lattice sites occupied by spins with magnitudes $S$ and $\tau$; 
we denote the set of magnitudes by ($S$, $\tau$). 
One of the authors and coworkers\cite{takano,Takano-K-S} introduced this lattice structure and  generally investigated the case of ($S$, $S$), i.e., the pure diamond chain (PDC). After that,  Niggemann et al.\cite{nig1,nig2} argued about a series of diamond chains with ($S$, 1/2).

%====================================
%====================================

The mixed diamond chain (MDC) is defined as a %the
 diamond chain with ($S$, $S/2$) for the integer $S$.\cite{tsh} Recently, extensive investigation on the MDC has been carried out by the present authors.\cite{tsh,hts,htsalt,htsus} The MDC is of special interest among diamond chains, because only the MDC has the Haldane phase in the absence of  frustration   so that we can observe the transition from the Haldane phase to a SCS phase induced by frustration. In contrast, diamond chains of  other types have  ferrimagnetic ground states  for weak frustration.

The common features of any types of diamond chains are that they have an infinite number of local conservation laws and more than two different types of exact SCS ground states are realized depending on the strength of frustration. The MDC with (1,$1/2$)  has  3 different paramagnetic SCS phases accompanied by spontaneous translational symmetry breakdown (STSB) and  one paramagnetic phase without STSB in addition to a nonmagnetic Haldane ground state in a less frustrated region.\cite{nig1, tsh} The SCS structures of the ground states are also reflected in characteristic thermal properties, as reported in ref. \citen{hts}. 

Thus, the diamond chain serves as a playground for investigating  various additional effects on frustrated magnetism on a well-founded basis. Therefore, various modifications of the PDC and MDC have been examined by  many researchers. 
\cite{ottk,otk,sano,dia4spin,htsalt,htsus} It is found that the natural mineral azurite consists of distorted PDCs with spin 1/2 and the magnetic properties of this material have been experimentally studied in detail.\cite{kiku,ohta} Other materials have also been reported.\cite{izuoka,uedia} In addition, as reviewed in ref. \citen{htsalt}, the MDC is related to many other important models of frustrated magnetism.\cite{plaq,plaq2,koga1,koga2,plaq3,plaq4,frulad1,str1,str2,fuku}

In the present work, we concentrate on the effects of the single-site anisotropy $D$ on the ground-state and finite-temperature properties of the MDC with (1,1/2). In many quantum spin systems, it is known that the $D$-term changes the ground state drastically. A well-known example is the effect of the $D$-term on the Haldane state of  spin-1 antiferromagnetic Heisenberg chains. A large positive $D$ term destroys the hidden order in the Haldane phase and drives the ground state into the large-$D$ phase that has no specific order. On the other hand,  a large negative $D$-term stabilizes the N\'eel order.
\cite{schulz, den-nijs, ht, chen,kt,oshikawa92} The $D$-terms in the higher spin and mixed spin Heisenberg chains induce rich phase diagrams.\cite{oshikawa92,tone_lsld,tone_dd,tone_dd_pla}  The alternating $D$-terms in  spin-1 
Heisenberg chains can pin the fluctuating hidden order to induce a long-period N\'eel order.\cite{hc} They can also induce the partial ferrimagnetic order in higher-spin chains.\cite{kh} Considering the high degeneracy of the ground states of the MDC\cite{tsh}, we may expect an even richer phase diagram in the MDC with the single-site anisotropy.

Thus far, despite the theoretical relevance of the MDC, no materials described by the MDC have been found. Nevertheless, synthesizing MDC materials is not an unrealistic expectation in view of the successful  synthesis of many novel magnetic materials such as molecule-based magnetic materials.\cite{m-d}
In general, it is natural to expect  the single-site anisotropy on the spin-1 site in real MDC compounds. From this viewpoint, it is important to present theoretical predictions on the ground state and finite-temperature properties of anisotropic MDCs to widen the range of candidate materials of the MDC and raise the possibility of their synthesis. We find that the effect of the single-site anisotropy not only gives quantitative correction to  physical quantities but also changes the ground state and low-temperature behavior qualitatively. Among them, the SCS ground states are found to be sensitive to such anisotropy.

This paper is organized as follows. 
In \S 2, the Hamiltonian for the MDC with the single-site anisotropy is presented. 
In \S 3, the MDC with the single-site anisotropy is examined by numerical methods and the ground-state phase diagram is obtained. 
Various limiting cases in the phase diagram are discussed by the perturbation
method. 
In \S 4, the finite-temperature behaviors of the longitudinal and transverse susceptibilities and entropy are investigated numerically and analytically by extending 
 the method developed in ref. \citen{hts}.   
The last  section is devoted to the summary and discussion.

\section{Hamiltonian}
We consider the MDC with the single-site anisotropy on spin-1 sites described by  the Hamiltonian
%------------------------------------------------------------
\begin{align}
{\cal H} =& {J} \sum_{l=1}^{L} \left[ 
(\v{S}_{l}+\v{S}_{l+1})\cdot(\v{\tau}^{(1)}_{l}+\v{\tau}^{(2)}_{l})+ \lambda\v{\tau}^{(1)}_{l}\v{\tau}^{(2)}_{l}\right]
\nonumber\\
&+D\sum_{l=1}^{L}S^{z2}_l , 
\label{hama}
\end{align}
%------------------------------------------------------------
where $\v{S}_{l}$ is the spin-1 operator, and 
$\v{\tau}^{(1)}_{l}$ and $\v{\tau}^{(2)}_{l}$ 
are the spin-1/2 operators in the $l$th unit cell. The total number of unit cells is denoted by $L$. 
In the case of $D = 0$, eq.~(\ref{hama})  reduces to 
the Hamiltonian  of 
the isotropic MDC.\cite{tsh} In what follows, we set 
 the energy unit as %by
 $J=1$.

Before proceeding, we analyze the classical version of the quantum Hamiltonian 
 (\ref{hama}). 
We obtain  the classical ground state for any $\lambda$ and $D$ in Appendix. 
We have four classical phases separated by the phase boundaries %determined by
 $\lambda = 2$ and $D=0$. 
In the two phases of $\lambda > 2$, the ground-state spin configurations can be locally modified without an energy increase. 
This classical situation corresponds to the quantum situation that 
there are an infinite number of low-energy states that are transformed  
from each other by local modification. 
Such quasi-degenerate low-energy states may enhance quantum fluctuations to contribute to the appearance of exotic quantum states  for large $\lambda$ 
 in the quantum system even for finite $D$.

\section{Ground-State Phase Diagram}

\subsection{General formulation for 
phase boundaries}

The Hamiltonian %eq. 
(\ref{hama}) has a series of conservation laws. 
To see it, we rewrite eq. (\ref{hama}) 
 in the form 
%------------------------------------------------------------
\begin{align}
{\mathcal H} = 
\sum_{l=1}^{L} \left[{ } (\v{S}_{l}\v{T}_{l}+\v{T}_{l}\v{S}_{l+1})
 + \frac{\lambda}{2}\left(\v{T}^2_{l}-\frac{3}{2}\right)\right]+D\sum_{l=1}^{L}S^{z2}_l , 
\label{ham2}
\end{align}
%------------------------------------------------------------
where the composite spin operators $\v{T}_l$ are defined as 
%------------------------------------------------------------
\begin{align}
\v{T}_{l} \equiv \v{\tau}^{(1)}_{l}+\v{\tau}^{(2)}_{l} 
\quad (l = 1, 2, \cdots L). 
\end{align}
%------------------------------------------------------------
Then, it is  evident that 
%------------------------------------------------------------
\begin{align}
[\v{T}_l^2, {\mathcal H}] = 0 \quad (l = 1, 2, \cdots L). 
\end{align}
%------------------------------------------------------------
Thus, we have $L$ conserved quantities $\v{T}_l^2$ for all $l$, even for $D \neq 0$. 
By defining the magnitude $T_l$ of  $\v{T}_l$ by $\v{T}_l^2 = T_l (T_l + 1)$, we have  a 
 set of good quantum numbers $\{T_l; l=1,2,...,L\}$. 
Each $T_l$ takes a value of 0 or 1. 
The total Hilbert space of the Hamiltonian (\ref{ham2}) consists of 
separate subspaces, each of which is specified by 
a definite set of $\{T_l\}$, i.e., a sequence of 0 and 1.

A spin pair of $T_l=0$ is a singlet dimer, that cuts off the correlation between $\v{S}_{l}$'s at  
both sides as seen 
from eq.~(\ref{ham2}). 
Hence, when a segment is bounded by two $T_l=0$ pairs, it is isolated from 
 other parts of the spin chain. 
The segment including $n$ successive $\v{T}_{l}$'s  with $T_l=1$ and $n+1$ $\v{S}_{l}$'s 
 is called a cluster-$n$ as in the isotropic case\cite{tsh}. 
A 
 cluster-$n$ is equivalent to an antiferromagnetic Heisenberg chain consisting  of $2n+1$ effective spins with magnitude 1 
 with alternating anisotropy. 
The  Hamiltonian is written as
\begin{align}
{\mathcal {H}}_n = \sum_{l=1}^{2n} { } \tilde{\v{S}}_{l}\tilde{\v{S}}_{l+1}+D\sum_{l=1}^{n+1}\tilde{S}^{z2}_{2l-1} , 
\label{ham3}
\end{align}
where $\tilde{\v{S}}_{2l-1}={\v{S}}_{l}$ and $\tilde{\v{S}}_{2l}={\v{T}}_{l}$. 

The ground state of the total diamond chain is analyzed by describing each cluster by 
 the Heisenberg model (\ref{ham3}). 
For $D=0$, we have obtained the complete ground-state phase diagram 
in ref. \citen{tsh}. 
In particular, we found successive phase transitions between dimer-cluster-$n$ (DC$n$) phases with $n$ = 0, 1, 2, 3, and $\infty$ as $\lambda$ decreases.  
The DC0 phase is also called the dimer monomer phase where $T_l = 0$ for all $l$.  
The DC$n$ phase with $n$ = 1, 2, or 3 is the phase where the ground state  consists of an alternating array of cluster-$n$'s and singlet dimers. 
The DC$\infty$ phase is the Haldane phase for the equivalent Heisenberg chain with $T_l = 1$ for all $l$.  
The description in terms of cluster-$n$ and DC$n$ states is naturally extended to the finite-$D$ regimes.
By the same argument as that in ref. \citen{tsh}, the phase boundary between  the DC$(n-1)$ and  DC$n$ phases 
for finite $D$ is given by 
%------------------------------------------------------------
\begin{align} 
\lefteqn{\lambda_{\rm c}(n-1,n; D) }\nonumber\\
&= (n+1)\EHal_{\rm G}(2n-1,D)
-n\EHal_{\rm G}(2n+1,D), 
\label{bdry}
\end{align}
%------------------------------------------------------------
where $\EHal_{\rm G}(2n+1,D)$ is the ground-state energy of  the Hamiltonian %eq.
 (\ref{ham3}). These phase transitions are of the first order, since they take place as level crossings between two eigenstates of the original Hamiltonian %eq. 
(\ref{hama})
 characterized by different sets of quantum numbers $\{T_l\}$. 
The direct transition from the DC$n$ phase to the DC$\infty$ phase takes place at 
%------------------------------------------------------------
\begin{align}
\lambda_{\rm c}(n,\infty {; D})&=\EHal_{\rm G}(2n+1,D)- {2}(n+1)\epsgtilde(\infty,D), 
\label{bdryinf}
\end{align}
%------------------------------------------------------------
if $\lambda_{\rm c}(n,\infty {; D}) {>} \lambda_{\rm c}(n,n+1 {; D})$, where $\epsgtilde(\infty,D)$ is the ground-state energy of the Hamiltonian %eq.
 (\ref{ham3}) with $n \rightarrow \infty$ per unit cell. 

The above arguments are almost parallel to those in the isotropic case. For the physical conclusion, however, we observe a 
 significant difference from the isotropic case, as described in the following sections.

\subsection{Ground-state phase diagram determined by numerical method}
\begin{figure} 
%\centerline{\includegraphics[width=5.5cm]{lambdacall.eps}}
\centerline{\includegraphics[width=5.5cm]{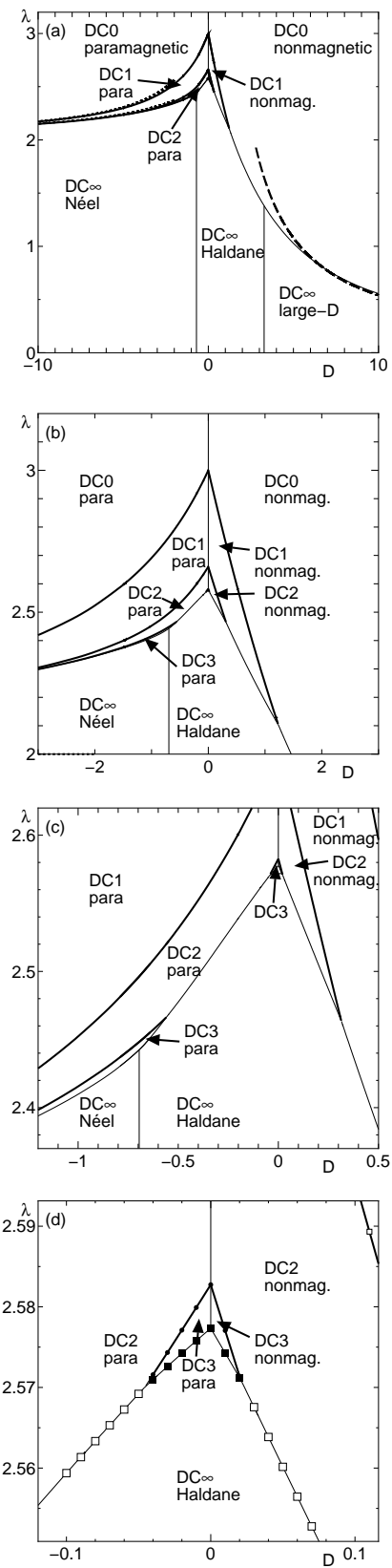}}
\caption{Ground-state phase diagram of the anisotropic mixed diamond chain. The figures are magnified appropriately from (a) to (d). In (a) and (b), some phases are not indicated because they are invisibly narrow on the scale of these figures. In (a), the analytical results in \S \ref{limits} are also shown 
by the dotted and broken lines.}
\label{phase}
\end{figure}

To obtain the ground-state phase diagram using eqs. (\ref{bdry}) and (\ref{bdryinf}), we have numerically diagonalized the Hamiltonian %eq.
 (\ref{ham3}) 
to calculate the ground-state energy $\EHal_{\rm G}(2n+1,D)$. The ground-state energy per unit cell $\epsgtilde(\infty,D)$ is calculated by the infinite-size DMRG by estimating the energy increment accompanied by an addition of a unit cell at the center of the chain. The phase diagram is shown in Fig. \ref{phase}. 
 Because {of} the presence of fine structures of the phase diagram, we present a series of phase diagrams magnified appropriately. In the top figure, the broken  and dotted lines are the results of the perturbation analysis described in the next section.

{The main features characteristic of the anisotropic case are as follows. 
\begin{enumerate}
\item The total spin $S_{\rm tot}$ of 
 cluster-$n$ is not a good quantum number, while its $z$-component remains a good quantum number. The ground state of %the
 cluster-$n$ has $S^{z}_{\rm tot}=0$ for $D>0$ and $S^{z}_{\rm tot}=\pm 1$ for $D < 0$. Therefore, the DC$n$ phase with finite $n$ is nonmagnetic for $D>0$ and paramagnetic for $D \leq  0$.
\item The DC$\infty$ state corresponds to the ground state of a spin-1 Heisenberg chain that has the single-site anisotropy 0 and $D$ alternatingly. It is known that the ground-state phase diagram of this model consists of a N\'eel phase for
$D \lesssim  -0.695$, a Haldane phase for $-0.695 \lesssim  D \lesssim  3.28$,
and a large-$D$ phase for $D \gtrsim 3.28$.\cite{hc}

\item  There appear DC$n$ phases with $n =$ 0, 1, 2, and 3 if $\lambda$ is above the critical value, which depends on $D$.  These phases continue to the isotropic DC$n$ phases. 
\item 
There appears another DC3 phase separated from the line of $D = 0$ and in contact with the end-point of the N\'eel-Haldane transition line in the DC$\infty$ region. 
\item The critical values of $\lambda$ between the DC$n$ phases decrease with $|D|$ from those in the isotropic case.  
\item For sufficiently large values of  
$|D|$, only the DC0 and DC$\infty$ phases exist. 
\end{enumerate}
}

\subsection{Limiting cases}
\label{limits}
\subsubsection{DC0-DC1 phase boundary}

 The phase boundary between the DC0 and DC1 phases is obtained by substituting the eigenvalues $\EHal_{\rm G}(1,D)$ and $\EHal_{\rm G}(3,D)$ into eq. (\ref{bdry}). These  eigenvalues can be analytically determined. 
Obviously,  {$\EHal_{\rm G}(1,D)=0$ for $D \geq 0$ and $\EHal_{\rm G}(1,D)=D$ for $D \leq 0$.} 

For $D > 0$, the ground state of a cluster-$n$ belongs to the subspace of 
$S^z_{\rm tot}=0$, $P=1$, and $T=1$, where $S^z_{\rm tot}$ is
the $z$-component of the total spin, $P$ is the space inversion parity,
and $T$ is the time reversal parity.
 Within this subspace, 
$\EHal_{\rm G}(3,D)$ is the smallest solution of 
the following eigenvalue equation: %------------------------------------------------------------
\begin{align}
&\EHal_{\rm G}(3,D)^3 
-
(3D-1)\EHal_{\rm G}(3,D)^2 \nonumber\\
&+2(D^2-D -3 )\EHal_{\rm G}(3,D)+8D =0 . 
\end{align}
%------------------------------------------------------------
Therefore, $\lambda_{\rm c}(0,1)=-\EHal_{\rm G}(3,D)$. 
%------------------------------------------------------------
For small values of  $D$, 
this implies $\lambda_{\rm c}(0,1) \simeq 3 -\frac{13}{15}D$. 

For $D < 0$, the ground state of a cluster-$n$ belongs to the subspace of %with the $z$ component of total spin 
$S^z_{\rm tot}=\pm 1$ and %space inversion parity 
$P=1$. Within this subspace, 
$\EHal_{\rm G}(3,D)$ is the smallest solution of 
the following eigenvalue equation: 
\begin{align}
&\EHal_{\rm G}(3,D)^4-(5D-2 )\EHal_{\rm G}(3,D)^3\nonumber\\
&\quad +(8D^2-6D -5 )\EHal_{\rm G}(3,D)^2\nonumber\\
&\quad +(-4D^3+4D^2 +14D -6 )\EHal_{\rm G}(3,D)\nonumber\\
&\quad -8D(D-1)=0 . 
\end{align}
%------------------------------------------------------------
Therefore, $\lambda_{\rm c}(0,1)=2D-\EHal_{\rm G}(3,D)$. 
For small values of  $|D|$, 
this implies $\lambda_{\rm c}(0,1) \simeq 3 +\frac{13}{30}D$.

\subsubsection{$D \rightarrow-\infty$}

For $D<0$ and $|D| \gg 1$, we apply the degenerate perturbation theory to the ground-state energy of a cluster-$n$. We regard the diagonal part of ${\mathcal H}_n $  as the unperturbed Hamiltonian ${\mathcal H}^{(0)}_n$ and the off-diagonal part as the perturbation Hamiltonian ${\mathcal H}^{(1)}_n$ as follows:
%------------------------------------------------------------
\begin{align}
{\mathcal H}^{(0)}_{n} &= D\sum_{l=1}^{n+1}S^{z2}_l + 
\sum_{l=1}^{n}   (S_l^z+S_{l+1}^z)T_{l}^z , 
\nonumber\\
{\mathcal H}^{(1)}_{n} &= 
{\frac{\alpha}{2}}
\sum_{l=1}^{n}  
\left[ (S_l^+ +S_{l+1}^+)T_{l}^- + \mathrm{h. \, c.} \right]. 
\label{ham22}
\end{align}
%------------------------------------------------------------
The expansion parameter $\alpha$ is introduced, which will be set equal to unity after the calculation. Up to the third-order perturbation calculation in $\alpha$, we have 
%------------------------------------------------------------
\begin{align}
&\EHal_{\rm G}(2n+1,D) {\simeq}
-(n+1)|D|-2n -\frac{(2n-2) }{|D|+3 }\alpha^2
\nonumber\\
&\qquad -\frac{2 }{|D|+2 }\alpha^2
-\frac{(3n-3) }{2(|D|+3 )^2}\alpha^3-\frac{1}{(|D|+2 )^2}\alpha^3 
\end{align}
%------------------------------------------------------------
for $n \geq 1$. 
Substituting this expression into eq. (\ref{bdry}),
the phase boundaries are determined up to the third order in $\alpha$ as follows. 
\begin{enumerate}
\item The DC$n$-DC$(n+1)$ transition point for $n \ge 1$ 
is given by 
%------------------------------------------------------------
\begin{align}
\lambda_c
&(n,n+1{; D})
\simeq 2 +\frac{4 }{|D|+3 }\alpha^2-\frac{2 }{|D|+2 }\alpha^2\nonumber\\
&\qquad +\frac{3 }{(|D|+3 )^2}\alpha^3-\frac{1}{(|D|+2 )^2}\alpha^3. \label{lmdc_third}
\end{align}
%------------------------------------------------------------
The $n$ independence of eq. (\ref{lmdc_third}) means the direct transition between the DC1 and  DC$\infty$ phases within this approximation. 
\item The DC0-DC1 transition point
 is given by 

%------------------------------------------------------------
\begin{align}
&\lambda_c
(0,1{; D}) 
\simeq 2\left(1 +\frac{1}{|D|+2 }\alpha^2
+\frac{1}{2(|D|+2 )^2}\alpha^3\right) , 
\end{align}
%------------------------------------------------------------
where the exact relation $\EHal_{\rm G}(1,D)=-|D|$ is used. 
\end{enumerate}
Setting $\alpha=1$,  these approximate phase boundaries 
 are drawn by the dotted lines in Fig. \ref{phase}(a).
\subsubsection{$D\rightarrow\infty$}

In this limit, all the $S$-spins are in the state $S^z=0$. 
Therefore, we find the following:

\begin{enumerate}
\item The ground-state energy of the DC$0$ phase is $\displaystyle-\frac{3L}{4}\lambda$. 
\item In the DC$\infty$ phase, the effective interaction between $T$-spins is $O(1/D)$ as described in ref. \citen{hc}. The effective Hamiltonian is given by
%------------------------------------------------------------
\begin{align}
&{\cal{H}}_{\rm eff} =\frac{L}{4}\lambda\nonumber\\
&\quad +\sum_{l=1}^L\left[\Jeff(T^x_lT^x_{l+1}+T^y_lT^y_{l+1})+\Deff(T^{z2}_l-2)\right],\\
&\Jeff =\Deff=\frac{2 }{D}.
\end{align}
%------------------------------------------------------------
The ground-state energy of this effective model is calculated by the infinite-size DMRG method as $\displaystyle\frac{\lambda}{4}-2.6995\Jeff$ per site.
According to the numerical calculation, there appear no DC$n$ phases with $1 \leq n < \infty$ for $D > 1.24 $. 
Taking this into account, we conclude that the direct transition between DC0 and DC$\infty$ phases takes place at 
%------------------------------------------------------------
\begin{align}
\lambda_{\rm c}(0,\infty {; D})\simeq \frac{5.399}{D}
\end{align}
%------------------------------------------------------------
for sufficiently large values of 
 $D$ up to $O(1/D)$.
This approximate phase boundary 
 is drawn by the broken line in Fig. \ref{phase}(a).
\end{enumerate}

\section{Finite-Temperature Properties}
\subsection{General formula}
Following the method described in ref. \citen{hts}, we can calculate the thermodynamic expectation value of the extensive  quantity $Q$  per unit cell as
%------------------------------------------------------------
\begin{align}
\frac{\aver{Q}}{L}=\frac{\aver{Q_{\rm cl}}}{\aver{L_{\rm cl}}} , 
\label{genq}
\end{align}
%------------------------------------------------------------
where $Q_{\rm cl}$ is the physical quantity $Q$ for each cluster-$n$ and ${L_{\rm cl}}\equiv n+1$ is the  number of unit cells per cluster-$n$.  
The average $\aver{Q_{\rm cl}}$ of  $Q_{\rm cl}$  over the grand canonical ensemble is 
written as 
%------------------------------------------------------------
\begin{align}
\aver{Q_{\rm cl}}&=\frac{\displaystyle\sum_{n=0}^{\infty}Z_{\rm cl}(n)\aver{Q_{\rm cl}(n)}_{\rm can}e^{\beta(\mu-\lambda/4) (n+1)}}{\displaystyle\sum_{n=0}^{\infty}Z_{\rm cl}(n) e^{\beta(\mu-\lambda/4) (n+1)}} , 
\label{acl}
\end{align}
%------------------------------------------------------------
where $\mu$ is the chemical potential of a cluster-$n$. 
$\aver{Q_{\rm cl}(n)}_{\rm can}$ and $Z_{\rm cl}(n)$ are the canonical average of the physical quantity $Q$ and the partition function of a cluster-$n$ with fixed $n$, respectively. These are defined by 
%------------------------------------------------------------
\begin{align}
\aver{Q_{\rm cl}(n)}_{\rm can}
&=\frac{1}{Z_{\rm cl}(n)}\sum_{\scl^z}\sum_{\nu}e^{-\beta \EHal(2n+1,{\nu};\scl^z)} \nonumber\\
\times &\bra{2n+1,\nu;\scl^z}Q_{\rm cl}\ket{2n+1,\nu;\scl^z} , \\
Z_{\rm cl}(n)&=\sum_{\scl^z}\sum_{\nu}e^{-\beta \EHal(2n+1,\nu;\scl^z)} , 
\label{cano}
\end{align}
%------------------------------------------------------------
where $\EHal(2n+1,\nu;\scl^z)$ and $\ket{2n+1,\nu;\scl^z}$ are, respectively, the $\nu$-th eigenenergy and eigenstate of the spin-1 chain with length $2n+1$ and magnetization $\scl^z$. 
The chemical potential $\mu$ is determined by 
the condition
%------------------------------------------------------------
\begin{align}
\sum_{n=0}^{\infty} e^{\beta\lambda} {Z}_{\rm cl}(n)e^{\beta(\mu-\lambda/4)(n+1)}=1 . 
\end{align}
%------------------------------------------------------------

\subsection{Formulae for entropy and magnetic susceptibility}

The entropy per unit cell is 
calculated using the same formula as that used in the isotropic case derived in ref. \citen{hts} as
%------------------------------------------------------------
\begin{align}
{\mathcal{S}} 
&=\frac{1}{T} \left(\frac{\aver{\Ecl}}{\aver{L_{\rm cl}}} - \mu \right) . 
\label{entform}
\end{align}
%-----------------
where $\Ecl=\EHal+(n-3)\lambda/4$ is the energy per cluster-$n$.
%-------------------------------------------

The longitudinal magnetic susceptibility $\chi_{\parallel}$ is 
calculated using the same formula as that used in the isotropic case by the direct application of eqs. (\ref{genq}), (\ref{acl}), and (\ref{cano}). This gives
%------------------------------------------------------------
\begin{align}
\chi_{\parallel}&=\frac{1}{T}\frac{\aver{M_{\rm cl}^2}}{\aver{L_{\rm cl}}} . 
\end{align}
%------------------------------------------------------------
The transverse susceptibility $\chi_{\perp}$ is 
calculated as discussed below. 

The contribution to the Hamiltonian due to the transverse field is given by
%------------------------------------------------------------
\begin{align}
{\mathcal H}_x &=-H\sum_{l=1}^{L} ({S}^x_{l}+{T}^x_{l}) . 
\end{align}
%------------------------------------------------------------
Because each $\v{T}_l^2$ commutes with ${\mathcal H}_x$, all $T_l$'s are conserved even in the presence of the transverse field. Therefore, the magnetic susceptibility of an anisotropic MDC (\ref{hama}) is the sum of the contributions of cluster-$n$'s.

Using the standard linear response theory, the canonical expectation value of the transverse magnetic susceptibility of a cluster-$n$ is given by
%------------------------------------------------------------
\begin{align}
&\aver{\chi_{\rm cl \perp}(n)}_{\rm can}
=\frac{1}{Z_{\rm cl}(n)}
\sum_{\scl^z=0}^{2n}\sum_{{\nu_1},{\nu_2}}
\nonumber\\
&\quad\times\frac{e^{-\beta \EHal(2n+1,{\nu_1};\scl^z)}-e^{-\beta \EHal(2n+1,{\nu_2};\scl^z+1)}}{\EHal(2n+1,{\nu_2};\scl^z+1)-\EHal(2n+1,{\nu_1};\scl^z)}\nonumber\\
&\quad\times\left|\bra{2n+1,{\nu_2};\scl^z + 1}\scl^{+}\ket{2n+1,{\nu_1};\scl^z}\right|^2 , 
\label{sus_t}
\end{align}
%------------------------------------------------------------
where $\EHal(2n+1,\nu;\scl^z)$ and $\ket{2n+1,\nu;\scl^z}$ are the $\nu$-th eigenenergy and eigenstate of the spin-1 chain with length $2n+1$ and magnetization $\scl^z$.  Thus, the transverse susceptibility of the MDC per unit cell is given by
%------------------------------------------------------------
\begin{align}
\chi_{\perp}&=\frac{\aver{\chi_{\rm cl \perp}}}{\avercl{L}} , 
\end{align}
%------------------------------------------------------------
where $\aver{\cdots}$ means the 
grand canonical average defined in eq. (\ref{acl}). 

In the DC$n$ ground state, this formula  reduces to 
%------------------------------------------------------------
\begin{align}
\chi_{\perp}
&=\frac{1}{(n+1)}\sum_{\nu}\frac{1}{\EHal(2n+1,\nu;1)-\EHal_{\rm G}(2n+1,0)}
\nonumber\\
&\qquad\times\left|\bra{2n+1,{\nu};1}\scl^{+}\ket{2n+1,G;0}\right|^2\label{chiperp01}
\end{align}
%------------------------------------------------------------
for $D > 0$ and to
%------------------------------------------------------------
\begin{align}
\chi_{\perp}
=&\frac{1}{2(n+1)}\sum_{\nu,{\rm G}}\Big[\frac{1}{\EHal(2n+1,\nu;2)-\EHal_{\rm G}(2n+1,1)}\Big.
\nonumber\\
&\qquad\times\left|\bra{2n+1,{\nu};2}\scl^{+}\ket{2n+1,G;1}\right|^2
\nonumber\\
&+\frac{1}{\EHal(2n+1,\nu;0)-\EHal_{\rm G}(2n+1,-1)}
\nonumber\\
&\qquad\times\Big.\left|\bra{2n+1,{\nu};0}\scl^{+}\ket{2n+1,G;-1}\right|^2\Big]\label{chiperp02}
\end{align}
%------------------------------------------------------------
for $D < 0$. Here,  $\ket{2n+1,G;\scl}$ is the ground state of a cluster-$n$ with magnetization $\scl$.  For $D <0$, the ground states are degenerate, so that the summation is taken over all the ground states. It should be noted that $\chi_{\perp}$ remains finite at $T=0$ in both cases.

\subsection{Results for entropy and magnetic susceptibility}

In eq. (\ref{acl}), the summation over $n$ is 
taken over all non-negative integers. 
In the actual numerical calculation of finite-temperature properties, however, we can only include the contribution of cluster-$n$'s with finite $n$. Hence, we cannot expect  reliable results if the ground state is the DC$\infty$ state. Therefore, we limit ourselves to the parameter region with DC$n$ ground states with finite $n$. In what follows, we present the results for the magnetic susceptibility and entropy calculated including the contribution of cluster-$n$'s with $0 \leq n \leq 5$. The error due to this cutoff procedure is estimated for the isotropic case in  ref. \citen{hts}. Because this estimation is based on an entropic argument in the high-temperature region, it is also valid in the present case. Thus, we expect that the missing entropy is within 3\% of the total entropy. 

\begin{figure}[!h]
%\centerline{\includegraphics[width=5.5cm]{chiparat_m001sel.eps}}
%\centerline{\includegraphics[width=5.5cm]{chiper_m001sel_b.eps}}
%\centerline{\includegraphics[width=5.5cm]{entr_m001sel_c.eps}}
\centerline{\includegraphics[width=5.5cm]{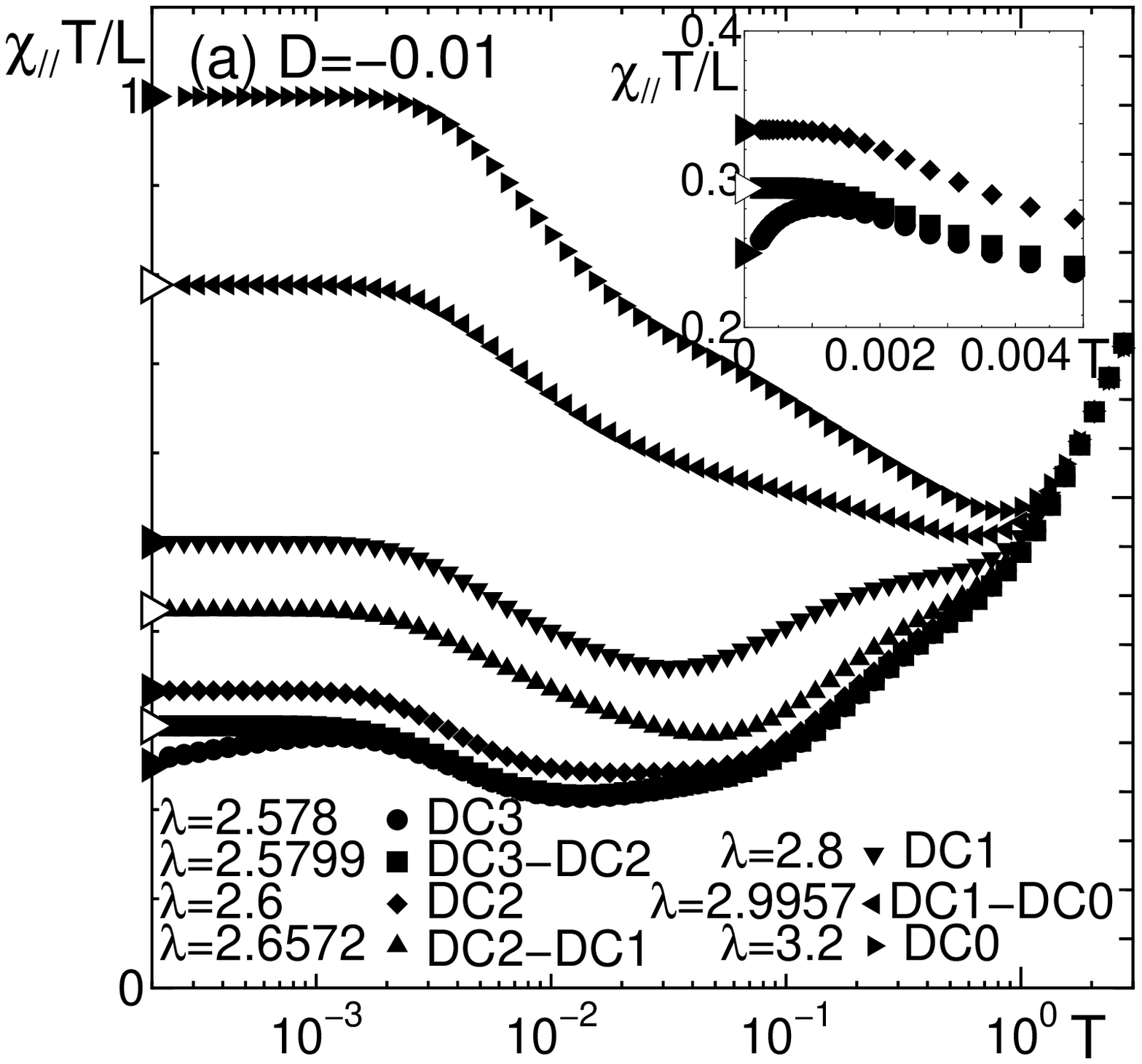}}
\centerline{\includegraphics[width=5.5cm]{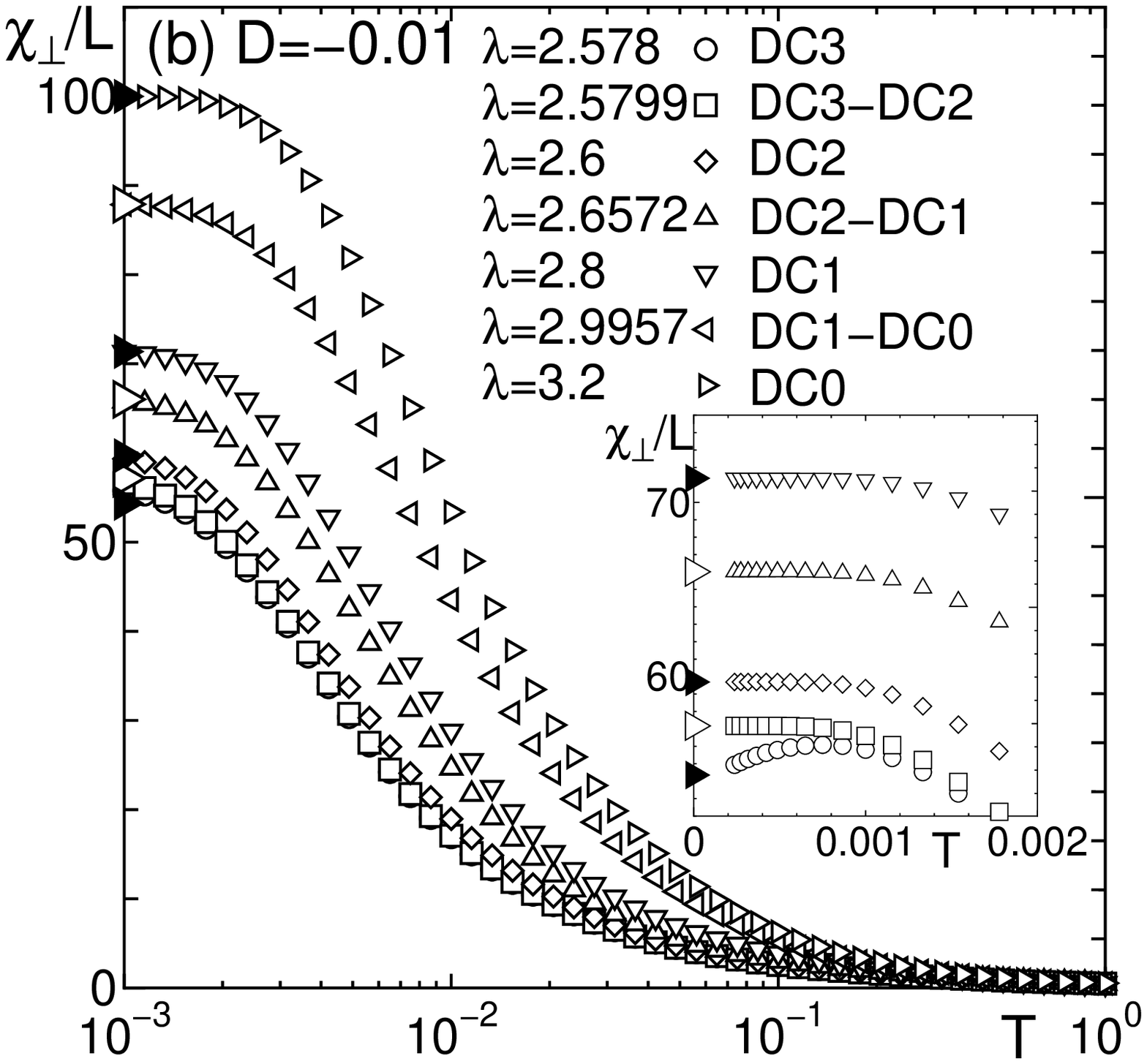}}
\centerline{\includegraphics[width=5.5cm]{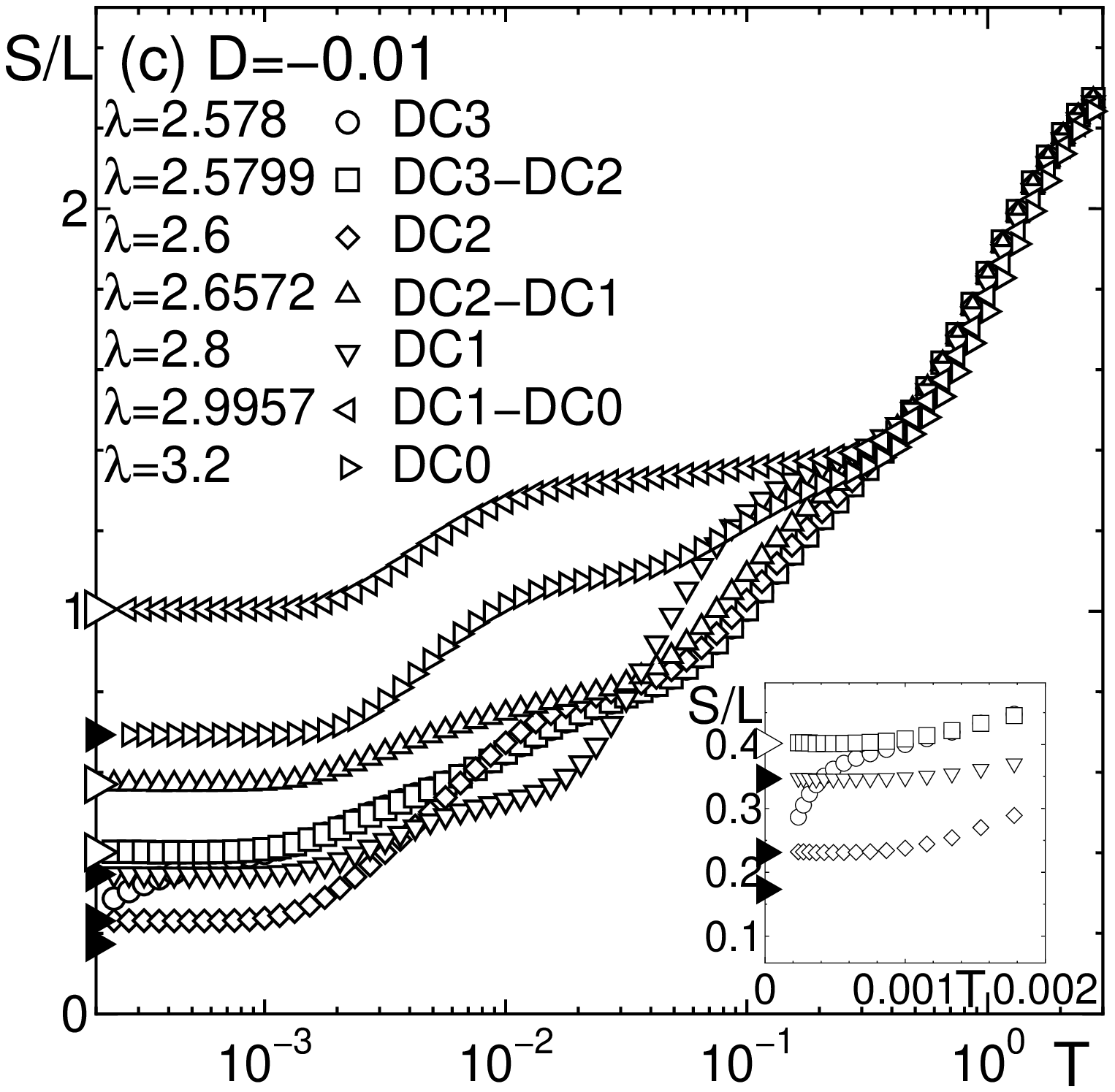}}
\caption{Temperature dependences of (a) longitudinal magnetic susceptibility $\chi_{\parallel}$ times $T$, (b) transverse  magnetic susceptibility $\chi_{\perp}$, and (c) entropy  $\mathcal S$ for $D=-0.01$. 
The chosen values of $\lambda$ are in the DC$n$ phases and also at the DC$(n+1)$-DC$n$ phase boundaries.  
The large filled right triangles on the ordinates  are the ground-state values in each phase and the large open ones are those at the phase boundaries.  
The insets are  magnified figures in the 
low-temperature regimes.}
\label{sust_para_m}
\end{figure}
\begin{figure}[!h]
%\centerline{\includegraphics[width=5.5cm]{chi001_sel.eps}}
%\centerline{\includegraphics[width=5.5cm]{entr_001sel_b.eps}}
\centerline{\includegraphics[width=5.5cm]{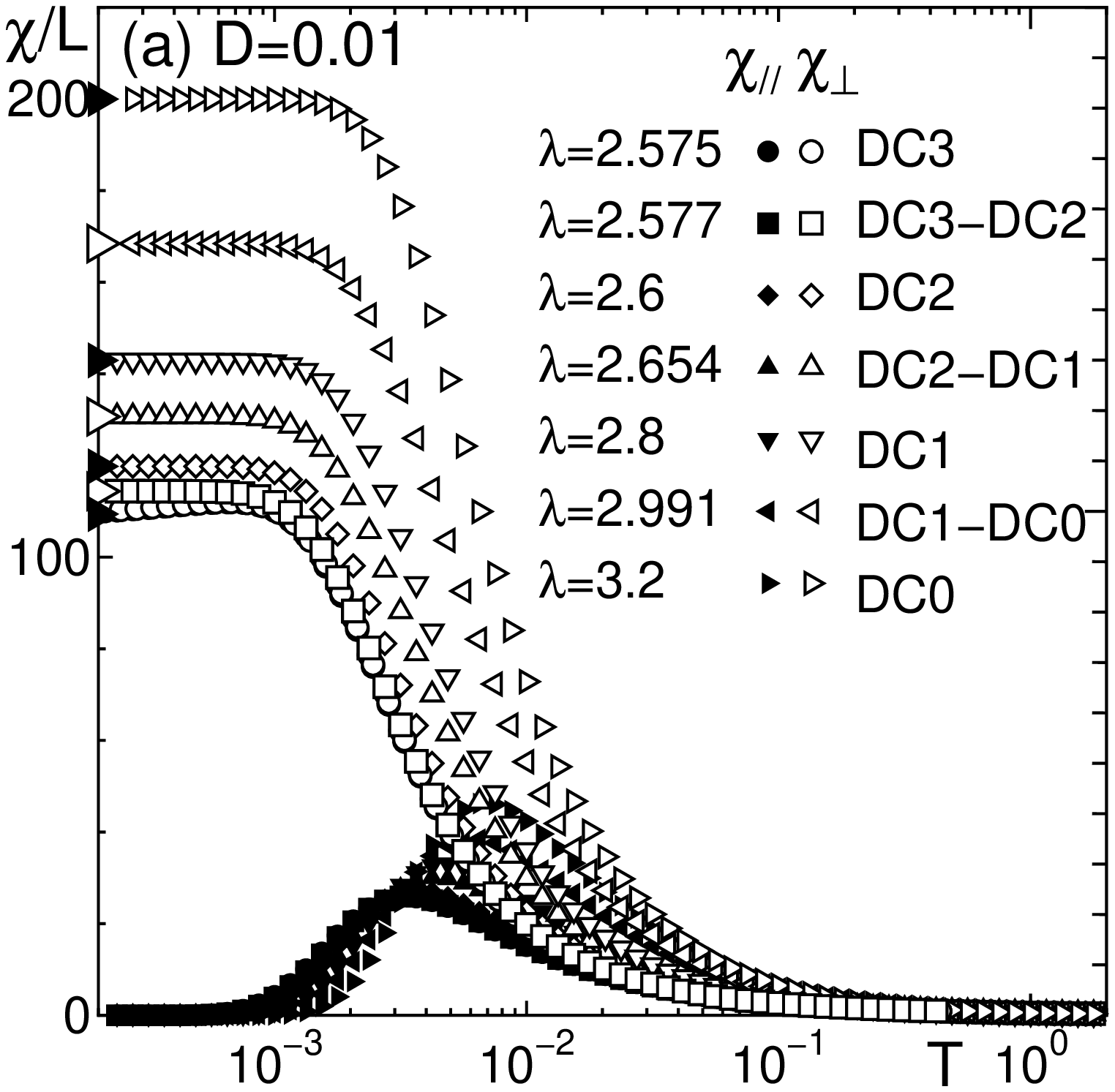}}
\centerline{\includegraphics[width=5.5cm]{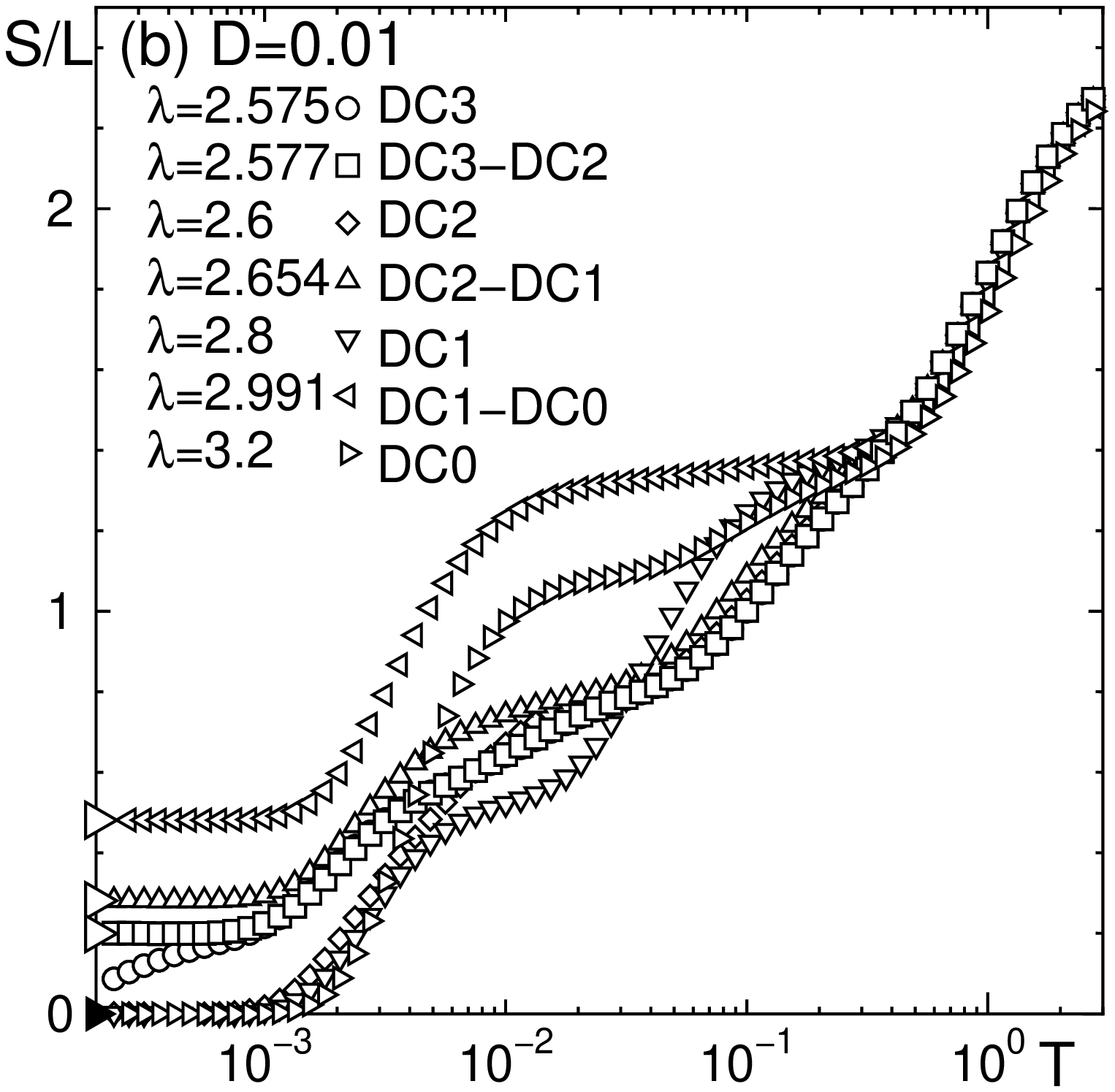}}
\caption{Temperature dependences of (a) longitudinal ($\chi_{\parallel}$) and transverse ($\chi_{\perp}$) magnetic susceptibilities and (b) entropy {$\mathcal{S}$} for 
 $D=0.01$
. 
 The chosen values of $\lambda$ are in the DC$n$ phases and also at DC$(n+1)$-DC$n$ phase boundaries.  
 The large filled right triangles on the ordinates  are the ground-state values in each phase and the large open ones are those at the phase boundaries. }
\label{sus_p}
\end{figure}

To demonstrate the sensitivity of the low-temperature behavior to the anisotropy $D$, we present the numerical results for $D = \pm 0.01$ as representatives of the easy-plane and easy-axis anisotropies. For a negative $D$, the longitudinal and transverse susceptibilities are shown in Figs. \ref{sust_para_m}(a) and \ref{sust_para_m}(b), respectively, and the entropy is shown in Fig. \ref{sust_para_m}(c). 
For a positive $D$, both the susceptibilities are shown in Fig. \ref{sus_p}(a) and the entropy is shown in Fig. \ref{sus_p}(b). 
 The difference between the low-temperature behaviors for the positive and negative $D$'s is distinct even for a small $|D|=0.01$. 
 In the low-temperature limit, the longitudinal magnetic susceptibility $\chi_{\parallel}$ and the entropy  $\mathcal{S}$ 
tend to zero for $D > 0$, except at the ground-state phase boundary. On the other hand, they behave as $\chi_{\parallel} T \simeq \dfrac{1}{n+1}$ and ${\mathcal{S}} \simeq \dfrac{\ln 2}{n+1}$ for $D < 0$ if the ground state belongs to the DC$n$ phase. 
The low-temperature limiting values of $\chi_{\perp}$ are calculated using eqs. (\ref{chiperp01}) and (\ref{chiperp02}). These limiting values 
 are shown by  the large filled  right triangles on the ordinates in  Figs. \ref{sust_para_m} and \ref{sus_p}.

 At the DC$(n-1)$-DC$n$ phase boundary, cluster-$(n-1)$ and cluster-$n$ coexist. In this case, the residual entropy can be estimated by the combinatory argument  similarly to that in the isotropic case. If we denote the number fraction of  cluster-$(n-1)$'s by $x$ and that of cluster-$n$'s by $1-x$, then the entropy can be estimated as
\begin{align}
{\mathcal{S}} 
&=-\frac{L}{n+1-x}\left[x\ln x +(1-x)\ln(1-x)-{\ln g}\right] , 
\end{align}
where $g$ is the degeneracy of the ground state of cluster-$n$, i.e., $g=1$ for $D >0$, $g=2$ for $D < 0$, and $g=3$ for $D=0$. By optimizing $\mathcal{S}$ 
with respect to $x$, we find
\begin{align}
 \frac{x}{g}&=\left(\frac{1-x}{x}\right)^n . 
\label{opt}
\end{align}
Using  $x$ that satisfies eq. (\ref{opt}), the entropy and susceptibility can be estimated as
\begin{align}
\frac{{\mathcal{S}}}{L}&=\ln \frac{x}{1-x} , \\
\frac{\chi_{\parallel,\perp}}{L}&=\frac{x\chi_{\rm cl \parallel,\perp}(n-1)+(1-x)\chi_{\rm cl \parallel,\perp}(n)}{nx+(n+1)(1-x)} . 
\end{align}
It should be noted that the residual entropy remains finite at the phase boundary even for $D >0$ owing to the mixing entropy of cluster-$(n-1)$ and cluster-$n$. 
For $D < 0$, $\chi_{\rm cl \parallel}=1/T$ for all $n$. Therefore, we have
\begin{align}
\frac{\chi_{\parallel}}{L}&=\frac{1}{T(n+1-x)} . 
\end{align}
For $D > 0$, $\dfrac{\chi_{\parallel}}{L}=0$. To calculate $\chi_{\perp}$, we have to estimate  $\chi_{\rm cl \parallel,\perp}(n-1)$ and $\chi_{\rm cl \parallel,\perp}(n)$ numerically using eqs. (\ref{chiperp01}) and  (\ref{chiperp02}). These low-temperature limiting values are shown by large open right triangles on the ordinates  in Figs.  \ref{sust_para_m} and \ref{sus_p}.

\section{Summary and Discussion}

We examined the anisotropic mixed diamond chain with spins 1 and 1/2 that has the single-site anisotropy $D$ on 
  spin-1 sites. 
In the ground-state phase diagram (Fig. \ref{phase}), 
there are  DC$n$ phases with $n =$ 0, 1, 2, and 3 if $\lambda$ is above the critical value, which depends on $D$.  
These phases continue to the isotropic DC$n$ phases 
except for a DC3 phase separated from the line of $D = 0$. 
As in the isotropic case, the DC$n$ ground state with finite $n$ is an alternating array of finite-length cluster-$n$'s and singlet dimers. 
It is nonmagnetic for the easy-plane anisotropy ($D {>} 0$), and is paramagnetic for the easy-axis anisotropy ($D {<} 0$). 
For smaller $\lambda$, the ground-state phase is one of the three DC$\infty$ phases:
i.e., the N\'eel phase for $D \lesssim  -0.695$, the Haldane phase for
$-0.695  \lesssim  D \lesssim  3.28$, and the large-$D$ phase for $D \gtrsim 3.28$.
The critical values of $\lambda$ between the DC$n$ phases decrease with $|D|$ from those in the isotropic case. For {sufficiently large values of} %enough
 $|D|$, only the DC0 and DC$\infty$ phases exist.  
The phase boundary between them tends to ${\lambda=}2$ for $D \rightarrow -\infty$ and to ${\lambda=}0$ for $D \rightarrow \infty$.

The above features of the ground-state phase diagram are 
qualitatively understood, if we incorporate
the effect of quantum fluctuation into 
 the classical  phase diagram explained in Appendix.
In the two classical phases with $\lambda > 2$, a pair of $\v{\tau}^{(1)}_{l}$ and $\v{\tau}^{(2)}_{l}$ can locally rotate around the fixed axis determined by $\v{S}_{l}$'s without increasing the energy, as seen in Appendix. 
The anisotropic term restricts the possible direction of the fixed axis: 
for $D<0$, the fixed axis is only parallel or antiparallel to the $z$-axis, 
and for $D>0$ it is confined in the $xy$ plane. 
The quantum effect of $D$ corresponding to this 
 classical restriction is to reduce the quantum fluctuation of $\v{S}_{l}$'s only. 
Therefore, for $\lambda \gtrsim 2$, the anisotropic term does not suppress the quantum fluctuation corresponding to the classical local rotations of $\v{\tau}^{(\nu)}_{l}$'s 
around the fixed axis of $\v{S}_{l}$. 
On the other hand, for $\lambda  \lesssim  2$, the classical ground state is antiferromagnetically ordered so that the quantum fluctuation 
is collectively concerned with all the spins 
and is not restricted  to individual pairs of $\v{\tau}^{(1)}_{l}$ and $\v{\tau}^{(2)}_{l}$. 
Therefore, the $D$-term suppresses the collective quantum fluctuation, and hinders the energy gain by the quantum fluctuation. 
This explains the reason why the region of the exotic ground state for large $\lambda$  extends to lower values of $\lambda$ with an increase in the magnitude of the anisotropy $|D|$.

The features of the ground-state phases and phase transitions are reflected 
 in physical quantities at low temperatures. 
The temperature dependences of entropy, and longitudinal and transverse susceptibilities are calculated. The low-temperature behavior is sensitive to the sign of $D$ for large $\lambda$, reflecting the change of the ground-state phase. The transverse susceptibility remains finite even at $T=0$, as long as $D \neq 0$. 

Generally, the introduction of anisotropy to a physical model increases the possibility that a corresponding material is realized. If an anisotropic MDC material is synthesized, the anisotropy may  
be controlled by changing the crystal field that reflects the environment of  spin-1 magnetic ions. This would be accomplished by, e.g., applying  pressure and/or changing the nonmagnetic ions around the magnetic ones. In contrast to the spin-1 chain with the Haldane gap, the ground state of the MDC is sensitive to the weak anisotropy. 
Therefore, 
 the paramagnetic-nonmagnetic transition should 
 take place with a small variation in the crystal field. 
It would also be possible to 
observe the quantum phase transitions between the DC$n$ states with different spatial periodicities.

In refs. \citen{htsalt} and \citen{htsus}, we have reported the effect of distortion on the isotropic MDC. It is predicted that various types of distortion can induce various types of quantum phases such as the Haldane phase with STSB, and the quantized and partial ferrimagnetic phases. The interplay of anisotropy and distortion would lead to an even wider variety of phenomena. Their investigation is left for future studies.

The numerical diagonalization program is based on the package TITPACK ver.2 coded by H. Nishimori.  The numerical computation in this work has been carried out using the facilities of the Supercomputer Center, Institute for Solid State Physics, University of Tokyo, the Supercomputing Division, Information Technology Center, University of Tokyo, and the  Yukawa Institute Computer Facility, Kyoto University.  KH is  supported by a Grant-in-Aid for Scientific Research (C) (21540379) from the Japan Society for the Promotion of Science.

\section*{Appendix}

We examine the phase diagram of the classical version of the quantum Hamiltonian %eq.
 (\ref{ham2}), where 
$\v{S}_{l}$ and $\v{\tau}_{l}^{(\nu)}$ are interpreted as classical vectors with  length $S$ and $\tau = S/2$, respectively. 
The classical Hamiltonian is then expressed in the following two forms: 
%------------------------------------------------------------
\begin{align} %
{\cal H}^{\mathrm{{clas}}} 
&= \frac{1}{4} \sum_l 
\left[ (2{\v{T}}_{l}+{\hat{\v{S}}}_{l})^{2} 
- 2(2 -\lambda) {\v{T}}_{l}^{2} 
- {{\hat{\v{S}}}}_{l}^{2} 
\right] 
+ {\cal H}^{\mathrm{{clas}}}_1 
\label{classic1}\\
&= \frac{1}{4\lambda} \sum_l 
\left[2(\lambda{\v{T}}_{l}+{\hat{\v{S}}}_{l})^{2} 
- 2{\hat{\v{S}}}_{l}^{2} 
\right] 
+  {\cal H}^{\mathrm{{clas}}}_1 , 
\label{classic2}
\end{align}
%------------------------------------------------------------
where ${\hat{\v{S}}}_{l} \equiv {\v{S}}_{l} + {\v{S}}_{l+1}$, 
$\v{T}_{l} \equiv \v{\tau}^{(1)}_{l}+\v{\tau}^{(2)}_{l}$, and 
{${\cal H}^{\mathrm{{clas}}}_1 \equiv \sum_{l=1}^{L} (DS^{z2}_l - \lambda S^2/4)$.}

In the case of $\lambda \le 2$, the expression %eq. 
(\ref{classic1}) shows that 
${\cal H}^{\mathrm{{clas}}}$ is minimized  
if $|{\hat{\v{S}}}_{l}|=2S$, $|{\v{T}}_{l}|=S$, 
and $|{\v{T}}_{l}+\frac{1}{2}{\hat{\v{S}}}_{l}|=0$ 
irrespective of $D$. 
Hence, all the ${\v{S}}_{l}$'s ($\v{\tau}_{l}^{(\alpha)}$'s) 
in the chain are aligned parallel (antiparallel) to a fixed axis, 
and the ground state is antiferromagnetic. 
The fixed axis is the $z$-axis for $D<0$, 
while it may be any direction in the $xy$ plane for $D>0$. 
The ground state is elastic in both cases, since any local modification of the spin configuration increases the energy.

In the case of $\lambda > 2$, the expression %eq.
 (\ref{classic2}) reveals that 
${\cal H}^{\mathrm{{clas}}}$ is minimized if $|{\hat{\v{S}}}_{l}|=2S$ 
and $|{\v{T}}_{l}+{\hat{\v{S}}}_{l}/\lambda |=0$ irrespective of $D$. 
Hence, ${\v{S}}_{l}$'s in the chain are aligned parallel to a fixed axis, 
and $\v{\tau}_{l}^{(1)}$ and $\v{\tau}_{l}^{(2)}$ 
form a triangle with ${\hat{\v{S}}}_{l}/\lambda$. 
The fixed axis is the $z$-axis for $D<0$, 
while it may be any direction in the $xy$ plane for $D>0$. 
Then, the ground state has a local arbitrariness, because 
$\v{\tau}_{l}^{(1)}$ and $\v{\tau}_{l}^{(2)}$ may be rotated about the axis of ${\v{S}}_{l}$ and ${\v{S}}_{i+1}$ without raising the energy. 
All the ground states are ferrimagnetic with a magnetization $(1 - 2/\lambda)SN$.

Thus, we have the classical phase diagram consisting of the four phases
separated by the phase boundaries of $\lambda=2$ and $D=0$.

\end{document}